\documentstyle[aps,multicol,preprint]{revtex}

\newcommand{\pr}{\partial}
\newcommand{\ub}{\overline{u}}
\newcommand{\baru}{\overline{u}}

\newcommand{\pb}{\overline{p}}
\newcommand{\uub}{{\overline{{\bf{u}}}}}
\newcommand{\uu}{{\bf{u}}}
\newcommand{\be}{\begin{equation}}
\newcommand{\en}{\end{equation}}

\newcommand{\ovl}{\overline}
\newcommand{\bea}{\begin{eqnarray}}
\newcommand{\ena}{\end{eqnarray}}

\begin{document}

\def\PsfigVersion{1.10}
\def\setDriver{\DvipsDriver} 
\ifx\undefined\psfig\else \fi
%

\let\LaTeXAtSign=\@
\let\@=\relax
\edef\psfigRestoreAt{\catcode`\@=\number\catcode`@\relax}
\catcode`\@=11\relax
\newwrite\@unused
\def\ps@typeout#1{{\let\protect\string\immediate\write\@unused{#1}}}

\def\DvipsDriver{
	\ps@typeout{psfig/tex \PsfigVersion -dvips}
\def\PsfigSpecials{\DvipsSpecials} 	\def\ps@dir{/}
\def\ps@predir{} }
\def\OzTeXDriver{
	\ps@typeout{psfig/tex \PsfigVersion -oztex}
	\def\PsfigSpecials{\OzTeXSpecials}
	\def\ps@dir{:}
	\def\ps@predir{:}
	\catcode`\^^J=5
}


\def\figurepath{./:}
\def\psfigurepath#1{\edef\figurepath{#1:}}

\def\DoPaths#1{\expandafter\EachPath#1\stoplist}
\def\leer{}
\def\EachPath#1:#2\stoplist{
  \ExistsFile{#1}{\SearchedFile}
  \ifx#2\leer
  \else
    \expandafter\EachPath#2\stoplist
  \fi}
%
%
\def\ps@dir{/}
\def\ExistsFile#1#2{%
   \openin1=\ps@predir#1\ps@dir#2
   \ifeof1
       \closein1
   \else
       \closein1
        \ifx\ps@founddir\leer
           \edef\ps@founddir{#1}
        \fi
   \fi}
%
%
\def\get@dir#1{%
  \def\ps@founddir{}
  \def\SearchedFile{#1}
  \DoPaths\figurepath
}

%
%
\def\@nnil{\@nil}
\def\@empty{}
\def\@psdonoop#1\@@#2#3{}
\def\@psdo#1:=#2\do#3{\edef\@psdotmp{#2}\ifx\@psdotmp\@empty \else
    \expandafter\@psdoloop#2,\@nil,\@nil\@@#1{#3}\fi}
\def\@psdoloop#1,#2,#3\@@#4#5{\def#4{#1}\ifx #4\@nnil \else
       #5\def#4{#2}\ifx #4\@nnil \else#5\@ipsdoloop #3\@@#4{#5}\fi\fi}
\def\@ipsdoloop#1,#2\@@#3#4{\def#3{#1}\ifx #3\@nnil 
       \let\@nextwhile=\@psdonoop \else
      #4\relax\let\@nextwhile=\@ipsdoloop\fi\@nextwhile#2\@@#3{#4}}
\def\@tpsdo#1:=#2\do#3{\xdef\@psdotmp{#2}\ifx\@psdotmp\@empty \else
    \@tpsdoloop#2\@nil\@nil\@@#1{#3}\fi}
\def\@tpsdoloop#1#2\@@#3#4{\def#3{#1}\ifx #3\@nnil 
       \let\@nextwhile=\@psdonoop \else
      #4\relax\let\@nextwhile=\@tpsdoloop\fi\@nextwhile#2\@@#3{#4}}
%
\ifx\undefined\fbox
\newdimen\fboxrule
\newdimen\fboxsep
\newdimen\ps@tempdima
\newbox\ps@tempboxa
\fboxsep = 3pt
\fboxrule = .4pt
\long\def\fbox#1{\leavevmode\setbox\ps@tempboxa\hbox{#1}\ps@tempdima\fboxrule
    \advance\ps@tempdima \fboxsep \advance\ps@tempdima \dp\ps@tempboxa
   \hbox{\lower \ps@tempdima\hbox
  {\vbox{\hrule height \fboxrule
          \hbox{\vrule width \fboxrule \hskip\fboxsep
          \vbox{\vskip\fboxsep \box\ps@tempboxa\vskip\fboxsep}\hskip 
                 \fboxsep\vrule width \fboxrule}
                 \hrule height \fboxrule}}}}
\fi
%
%
\newread\ps@stream
\newif\ifnot@eof       
\newif\if@noisy        
\newif\if@atend        
\newif\if@psfile       
%
%
{\catcode`\%=12\global\gdef\epsf@start{
\def\epsf@PS{PS}
\def\epsf@getbb#1{%
%
%
\openin\ps@stream=\ps@predir#1
\ifeof\ps@stream\ps@typeout{Error, File #1 not found}\else
%
%
   {\not@eoftrue \chardef\other=12
    \def\do##1{\catcode`##1=\other}\dospecials \catcode`\ =10
    \loop
       \if@psfile
	  \read\ps@stream to \epsf@fileline
       \else{
	  \obeyspaces
          \read\ps@stream to \epsf@tmp\global\let\epsf@fileline\epsf@tmp}
       \fi
       \ifeof\ps@stream\not@eoffalse\else
%
%
       \if@psfile\else
       \expandafter\epsf@test\epsf@fileline:. \\%
       \fi
%
%
          \expandafter\epsf@aux\epsf@fileline:. \\%
       \fi
   \ifnot@eof\repeat
   }\closein\ps@stream\fi}%
%
%
\long\def\epsf@test#1#2#3:#4\\{\def\epsf@testit{#1#2}
			\ifx\epsf@testit\epsf@start\else
\ps@typeout{Warning! File does not start with `\epsf@start'.  It may not be a PostScript file.}
			\fi
			\@psfiletrue} 
%
%
{\catcode`\%=12\global\let\epsf@percent=
%
%
%
\long\def\epsf@aux#1#2:#3\\{\ifx#1\epsf@percent
   \def\epsf@testit{#2}\ifx\epsf@testit\epsf@bblit
	\@atendfalse
        \epsf@atend #3 . \\%
	\if@atend	
	   \if@verbose{
		\ps@typeout{psfig: found `(atend)'; continuing search}
	   }\fi
        \else
        \epsf@grab #3 . . . \\%
        \not@eoffalse
        \global\no@bbfalse
        \fi
   \fi\fi}%
%
%
\def\epsf@grab #1 #2 #3 #4 #5\\{%
   \global\def\epsf@llx{#1}\ifx\epsf@llx\empty
      \epsf@grab #2 #3 #4 #5 .\\\else
   \global\def\epsf@lly{#2}%
   \global\def\epsf@urx{#3}\global\def\epsf@ury{#4}\fi}%
%
%
\def\epsf@atendlit{(atend)} 
\def\epsf@atend #1 #2 #3\\{%
   \def\epsf@tmp{#1}\ifx\epsf@tmp\empty
      \epsf@atend #2 #3 .\\\else
   \ifx\epsf@tmp\epsf@atendlit\@atendtrue\fi\fi}


\chardef\psletter = 11 
\chardef\other = 12

\newif \ifdebug 
\newif\ifc@mpute 
\c@mputetrue 

\let\then = \relax
\def\r@dian{pt }
\let\r@dians = \r@dian
\let\dimensionless@nit = \r@dian
\let\dimensionless@nits = \dimensionless@nit
\def\internal@nit{sp }
\let\internal@nits = \internal@nit
\newif\ifstillc@nverging
\def \Mess@ge #1{\ifdebug \then \message {#1} \fi}

{ 
	\catcode `\@ = \psletter
	\gdef \nodimen {\expandafter \n@dimen \the \dimen}
	\gdef \term #1 #2 #3%
	       {\edef \t@ {\the #1}
		\edef \t@@ {\expandafter \n@dimen \the #2\r@dian}%
		\t@rm {\t@} {\t@@} {#3}%
	       }
	\gdef \t@rm #1 #2 #3%
	       {{%
		\count 0 = 0
		\dimen 0 = 1 \dimensionless@nit
		\dimen 2 = #2\relax
		\Mess@ge {Calculating term #1 of \nodimen 2}%
		\loop
		\ifnum	\count 0 < #1
		\then	\advance \count 0 by 1
			\Mess@ge {Iteration \the \count 0 \space}%
			\Multiply \dimen 0 by {\dimen 2}%
			\Mess@ge {After multiplication, term = \nodimen 0}%
			\Divide \dimen 0 by {\count 0}%
			\Mess@ge {After division, term = \nodimen 0}%
		\repeat
		\Mess@ge {Final value for term #1 of 
				\nodimen 2 \space is \nodimen 0}%
		\xdef \Term {#3 = \nodimen 0 \r@dians}%
		\aftergroup \Term
	       }}
	\catcode `\p = \other
	\catcode `\t = \other
	\gdef \n@dimen #1pt{#1} 
}

\def \Divide #1by #2{\divide #1 by #2} 

\def \Multiply #1by #2
       {{
	\count 0 = #1\relax
	\count 2 = #2\relax
	\count 4 = 65536
	\Mess@ge {Before scaling, count 0 = \the \count 0 \space and
			count 2 = \the \count 2}%
	\ifnum	\count 0 > 32767 
	\then	\divide \count 0 by 4
		\divide \count 4 by 4
	\else	\ifnum	\count 0 < -32767
		\then	\divide \count 0 by 4
			\divide \count 4 by 4
		\else
		\fi
	\fi
	\ifnum	\count 2 > 32767 
	\then	\divide \count 2 by 4
		\divide \count 4 by 4
	\else	\ifnum	\count 2 < -32767
		\then	\divide \count 2 by 4
			\divide \count 4 by 4
		\else
		\fi
	\fi
	\multiply \count 0 by \count 2
	\divide \count 0 by \count 4
	\xdef \product {#1 = \the \count 0 \internal@nits}%
	\aftergroup \product
       }}

\def\r@duce{\ifdim\dimen0 > 90\r@dian \then   
		\multiply\dimen0 by -1
		\advance\dimen0 by 180\r@dian
		\r@duce
	    \else \ifdim\dimen0 < -90\r@dian \then  
		\advance\dimen0 by 360\r@dian
		\r@duce
		\fi
	    \fi}

\def\Sine#1%
       {{%
	\dimen 0 = #1 \r@dian
	\r@duce
	\ifdim\dimen0 = -90\r@dian \then
	   \dimen4 = -1\r@dian
	   \c@mputefalse
	\fi
	\ifdim\dimen0 = 90\r@dian \then
	   \dimen4 = 1\r@dian
	   \c@mputefalse
	\fi
	\ifdim\dimen0 = 0\r@dian \then
	   \dimen4 = 0\r@dian
	   \c@mputefalse
	\fi
	\ifc@mpute \then
		\divide\dimen0 by 180
		\dimen0=3.141592654\dimen0
		\dimen 2 = 3.1415926535897963\r@dian 
		\divide\dimen 2 by 2 
		\Mess@ge {Sin: calculating Sin of \nodimen 0}%
		\count 0 = 1 
		\dimen 2 = 1 \r@dian 
		\dimen 4 = 0 \r@dian 
		\loop
			\ifnum	\dimen 2 = 0 
			\then	\stillc@nvergingfalse 
			\else	\stillc@nvergingtrue
			\fi
			\ifstillc@nverging 
			\then	\term {\count 0} {\dimen 0} {\dimen 2}%
				\advance \count 0 by 2
				\count 2 = \count 0
				\divide \count 2 by 2
				\ifodd	\count 2 
				\then	\advance \dimen 4 by \dimen 2
				\else	\advance \dimen 4 by -\dimen 2
				\fi
		\repeat
	\fi		
			\xdef \sine {\nodimen 4}%
       }}

\def\Cosine#1{\ifx\sine\UnDefined\edef\Savesine{\relax}\else
		             \edef\Savesine{\sine}\fi
	{\dimen0=#1\r@dian\advance\dimen0 by 90\r@dian
	 \Sine{\nodimen 0}
	 \xdef\cosine{\sine}
	 \xdef\sine{\Savesine}}}	      

\def\psdraft{
	\def\@psdraft{0}
}
\def\psfull{
	\def\@psdraft{100}
}

\psfull

\newif\if@scalefirst
\def\psscalefirst{\@scalefirsttrue}
\def\psrotatefirst{\@scalefirstfalse}
\psrotatefirst

\newif\if@draftbox
\def\psnodraftbox{
	\@draftboxfalse
}
\def\psdraftbox{
	\@draftboxtrue
}
\@draftboxtrue

\newif\if@prologfile
\newif\if@postlogfile
\def\pssilent{
	\@noisyfalse
}
\def\psnoisy{
	\@noisytrue
}
\psnoisy
\newif\if@bbllx
\newif\if@bblly
\newif\if@bburx
\newif\if@bbury
\newif\if@height
\newif\if@width
\newif\if@rheight
\newif\if@rwidth
\newif\if@angle
\newif\if@clip
\newif\if@verbose
\def\@p@@sclip#1{\@cliptrue}
\newif\if@decmpr
\def\@p@@sfigure#1{\def\@p@sfile{null}\def\@p@sbbfile{null}\@decmprfalse
   \openin1=\ps@predir#1
   \ifeof1
	\closein1
	\get@dir{#1}
	\ifx\ps@founddir\leer
		\openin1=\ps@predir#1.bb
		\ifeof1
			\closein1
			\get@dir{#1.bb}
			\ifx\ps@founddir\leer
				\ps@typeout{Can't find #1 in \figurepath}
			\else
				\@decmprtrue
				\def\@p@sfile{\ps@founddir\ps@dir#1}
				\def\@p@sbbfile{\ps@founddir\ps@dir#1.bb}
			\fi
		\else
			\closein1
			\@decmprtrue
			\def\@p@sfile{#1}
			\def\@p@sbbfile{#1.bb}
		\fi
	\else
		\def\@p@sfile{\ps@founddir\ps@dir#1}
		\def\@p@sbbfile{\ps@founddir\ps@dir#1}
	\fi
   \else
	\closein1
	\def\@p@sfile{#1}
	\def\@p@sbbfile{#1}
   \fi
}
\def\@p@@sfile#1{\@p@@sfigure{#1}}
\def\@p@@sbbllx#1{
		\@bbllxtrue
		\dimen100=#1
		\edef\@p@sbbllx{\number\dimen100}
}
\def\@p@@sbblly#1{
		\@bbllytrue
		\dimen100=#1
		\edef\@p@sbblly{\number\dimen100}
}
\def\@p@@sbburx#1{
		\@bburxtrue
		\dimen100=#1
		\edef\@p@sbburx{\number\dimen100}
}
\def\@p@@sbbury#1{
		\@bburytrue
		\dimen100=#1
		\edef\@p@sbbury{\number\dimen100}
}
\def\@p@@sheight#1{
		\@heighttrue
		\dimen100=#1
   		\edef\@p@sheight{\number\dimen100}
}
\def\@p@@swidth#1{
		\@widthtrue
		\dimen100=#1
		\edef\@p@swidth{\number\dimen100}
}
\def\@p@@srheight#1{
		\@rheighttrue
		\dimen100=#1
		\edef\@p@srheight{\number\dimen100}
}
\def\@p@@srwidth#1{
		\@rwidthtrue
		\dimen100=#1
		\edef\@p@srwidth{\number\dimen100}
}
\def\@p@@sangle#1{
		\@angletrue
		\edef\@p@sangle{#1} 
}
\def\@p@@ssilent#1{ 
		\@verbosefalse
}
\def\@p@@sprolog#1{\@prologfiletrue\def\@prologfileval{#1}}
\def\@p@@spostlog#1{\@postlogfiletrue\def\@postlogfileval{#1}}
\def\@cs@name#1{\csname #1\endcsname}
\def\@setparms#1=#2,{\@cs@name{@p@@s#1}{#2}}
%
%
\def\ps@init@parms{
		\@bbllxfalse \@bbllyfalse
		\@bburxfalse \@bburyfalse
		\@heightfalse \@widthfalse
		\@rheightfalse \@rwidthfalse
		\def\@p@sbbllx{}\def\@p@sbblly{}
		\def\@p@sbburx{}\def\@p@sbbury{}
		\def\@p@sheight{}\def\@p@swidth{}
		\def\@p@srheight{}\def\@p@srwidth{}
		\def\@p@sangle{0}
		\def\@p@sfile{} \def\@p@sbbfile{}
		\def\@p@scost{10}
		\def\@sc{}
		\@prologfilefalse
		\@postlogfilefalse
		\@clipfalse
		\if@noisy
			\@verbosetrue
		\else
			\@verbosefalse
		\fi
}
%
%
\def\parse@ps@parms#1{
	 	\@psdo\@psfiga:=#1\do
		   {\expandafter\@setparms\@psfiga,}}
%
%
\newif\ifno@bb
\def\bb@missing{
	\if@verbose{
		\ps@typeout{psfig: searching \@p@sbbfile \space  for bounding box}
	}\fi
	\no@bbtrue
	\epsf@getbb{\@p@sbbfile}
        \ifno@bb \else \bb@cull\epsf@llx\epsf@lly\epsf@urx\epsf@ury\fi
}	
\def\bb@cull#1#2#3#4{
	\dimen100=#1 bp\edef\@p@sbbllx{\number\dimen100}
	\dimen100=#2 bp\edef\@p@sbblly{\number\dimen100}
	\dimen100=#3 bp\edef\@p@sbburx{\number\dimen100}
	\dimen100=#4 bp\edef\@p@sbbury{\number\dimen100}
	\no@bbfalse
}
\newdimen\p@intvaluex
\newdimen\p@intvaluey
\def\rotate@#1#2{{\dimen0=#1 sp\dimen1=#2 sp
		  \global\p@intvaluex=\cosine\dimen0
		  \dimen3=\sine\dimen1
		  \global\advance\p@intvaluex by -\dimen3
		  \global\p@intvaluey=\sine\dimen0
		  \dimen3=\cosine\dimen1
		  \global\advance\p@intvaluey by \dimen3
		  }}
\def\compute@bb{
		\no@bbfalse
		\if@bbllx \else \no@bbtrue \fi
		\if@bblly \else \no@bbtrue \fi
		\if@bburx \else \no@bbtrue \fi
		\if@bbury \else \no@bbtrue \fi
		\ifno@bb \bb@missing \fi
		\ifno@bb \ps@typeout{FATAL ERROR: no bb supplied or found}
			\no-bb-error
		\fi
		%
%
		\count203=\@p@sbburx
		\count204=\@p@sbbury
		\advance\count203 by -\@p@sbbllx
		\advance\count204 by -\@p@sbblly
		\edef\ps@bbw{\number\count203}
		\edef\ps@bbh{\number\count204}
		\if@angle 
			\Sine{\@p@sangle}\Cosine{\@p@sangle}
	        	{\dimen100=\maxdimen\xdef\r@p@sbbllx{\number\dimen100}
					    \xdef\r@p@sbblly{\number\dimen100}
			                    \xdef\r@p@sbburx{-\number\dimen100}
					    \xdef\r@p@sbbury{-\number\dimen100}}
%
                        \def\minmaxtest{
			   \ifnum\number\p@intvaluex<\r@p@sbbllx
			      \xdef\r@p@sbbllx{\number\p@intvaluex}\fi
			   \ifnum\number\p@intvaluex>\r@p@sbburx
			      \xdef\r@p@sbburx{\number\p@intvaluex}\fi
			   \ifnum\number\p@intvaluey<\r@p@sbblly
			      \xdef\r@p@sbblly{\number\p@intvaluey}\fi
			   \ifnum\number\p@intvaluey>\r@p@sbbury
			      \xdef\r@p@sbbury{\number\p@intvaluey}\fi
			   }
			\rotate@{\@p@sbbllx}{\@p@sbblly}
			\minmaxtest
			\rotate@{\@p@sbbllx}{\@p@sbbury}
			\minmaxtest
			\rotate@{\@p@sbburx}{\@p@sbblly}
			\minmaxtest
			\rotate@{\@p@sbburx}{\@p@sbbury}
			\minmaxtest
			\edef\@p@sbbllx{\r@p@sbbllx}\edef\@p@sbblly{\r@p@sbblly}
			\edef\@p@sbburx{\r@p@sbburx}\edef\@p@sbbury{\r@p@sbbury}
		\fi
		\count203=\@p@sbburx
		\count204=\@p@sbbury
		\advance\count203 by -\@p@sbbllx
		\advance\count204 by -\@p@sbblly
		\edef\@bbw{\number\count203}
		\edef\@bbh{\number\count204}
}
%
%
\def\in@hundreds#1#2#3{\count240=#2 \count241=#3
		     \count100=\count240	
		     \divide\count100 by \count241
		     \count101=\count100
		     \multiply\count101 by \count241
		     \advance\count240 by -\count101
		     \multiply\count240 by 10
		     \count101=\count240	
		     \divide\count101 by \count241
		     \count102=\count101
		     \multiply\count102 by \count241
		     \advance\count240 by -\count102
		     \multiply\count240 by 10
		     \count102=\count240	
		     \divide\count102 by \count241
		     \count200=#1\count205=0
		     \count201=\count200
			\multiply\count201 by \count100
		 	\advance\count205 by \count201
		     \count201=\count200
			\divide\count201 by 10
			\multiply\count201 by \count101
			\advance\count205 by \count201
		     \count201=\count200
			\divide\count201 by 100
			\multiply\count201 by \count102
			\advance\count205 by \count201
		     \edef\@result{\number\count205}
}
\def\compute@wfromh{
		\in@hundreds{\@p@sheight}{\@bbw}{\@bbh}
		\edef\@p@swidth{\@result}
}
\def\compute@hfromw{
	        \in@hundreds{\@p@swidth}{\@bbh}{\@bbw}
		\edef\@p@sheight{\@result}
}
\def\compute@handw{
		\if@height 
			\if@width
			\else
				\compute@wfromh
			\fi
		\else 
			\if@width
				\compute@hfromw
			\else
				\edef\@p@sheight{\@bbh}
				\edef\@p@swidth{\@bbw}
			\fi
		\fi
}
\def\compute@resv{
		\if@rheight \else \edef\@p@srheight{\@p@sheight} \fi
		\if@rwidth \else \edef\@p@srwidth{\@p@swidth} \fi
}
%
\def\compute@sizes{
	\compute@bb
	\if@scalefirst\if@angle
	\if@width
	   \in@hundreds{\@p@swidth}{\@bbw}{\ps@bbw}
	   \edef\@p@swidth{\@result}
	\fi
	\if@height
	   \in@hundreds{\@p@sheight}{\@bbh}{\ps@bbh}
	   \edef\@p@sheight{\@result}
	\fi
	\fi\fi
	\compute@handw
	\compute@resv}
\def\OzTeXSpecials{
	\special{empty.ps /@isp {true} def}
	\special{empty.ps \@p@swidth \space \@p@sheight \space
			\@p@sbbllx \space \@p@sbblly \space
			\@p@sbburx \space \@p@sbbury \space
			startTexFig \space }
	\if@clip{
		\if@verbose{
			\ps@typeout{(clip)}
		}\fi
		\special{empty.ps doclip \space }
	}\fi
	\if@angle{
		\if@verbose{
			\ps@typeout{(rotate)}
		}\fi
		\special {empty.ps \@p@sangle \space rotate \space} 
	}\fi
	\if@prologfile
	    \special{\@prologfileval \space } \fi
	\if@decmpr{
		\if@verbose{
			\ps@typeout{psfig: Compression not available
			in OzTeX version \space }
		}\fi
	}\else{
		\if@verbose{
			\ps@typeout{psfig: including \@p@sfile \space }
		}\fi
		\special{epsf=\ps@predir\@p@sfile \space }
	}\fi
	\if@postlogfile
	    \special{\@postlogfileval \space } \fi
	\special{empty.ps /@isp {false} def}
}
\def\DvipsSpecials{
	\special{ps::[begin] 	\@p@swidth \space \@p@sheight \space
			\@p@sbbllx \space \@p@sbblly \space
			\@p@sbburx \space \@p@sbbury \space
			startTexFig \space }
	\if@clip{
		\if@verbose{
			\ps@typeout{(clip)}
		}\fi
		\special{ps:: doclip \space }
	}\fi
	\if@angle
		\if@verbose{
			\ps@typeout{(clip)}
		}\fi
		\special {ps:: \@p@sangle \space rotate \space} 
	\fi
	\if@prologfile
	    \special{ps: plotfile \@prologfileval \space } \fi
	\if@decmpr{
		\openin1=\ps@predir\@p@sfile.gz
		\ifeof1
		        \closein1
			\if@verbose{
				\ps@typeout{psfig: including \@p@sfile.Z \space }
			}\fi
			\special{ps: plotfile "`zcat \@p@sfile.Z" \space }
		\else
                        \closein1
			\if@verbose{
				\ps@typeout{psfig: including \@p@sfile.gz \space }
			}\fi
			\special{ps: plotfile "`gunzip -c \@p@sfile.gz" \space }
		\fi
	}\else{
		\if@verbose{
			\ps@typeout{psfig: including \@p@sfile \space }
		}\fi
		\special{ps: plotfile \@p@sfile \space }
	}\fi
	\if@postlogfile
	    \special{ps: plotfile \@postlogfileval \space } \fi
	\special{ps::[end] endTexFig \space }
}
%
%
\def\psfig#1{\vbox {
	%
	\ps@init@parms
	\parse@ps@parms{#1}
	\compute@sizes
	\ifnum\@p@scost<\@psdraft{
		\PsfigSpecials 
		\vbox to \@p@srheight sp{
			\hbox to \@p@srwidth sp{
				\hss
			}
		\vss
		}
	}\else{
		\if@draftbox{		
			\hbox{\fbox{\vbox to \@p@srheight sp{
			\vss
			\hbox to \@p@srwidth sp{ \hss 
			 \hss }
			\vss
			}}}
		}\else{
			\vbox to \@p@srheight sp{
			\vss
			\hbox to \@p@srwidth sp{\hss}
			\vss
			}
		}\fi

	}\fi
}}
\psfigRestoreAt
\setDriver
\let\@=\LaTeXAtSign

\draft
\title{Commutator-errors in large-eddy simulation}

\author{Bernard J. Geurts}
\address{Mathematical Sciences, J.M. Burgers Center,
University of Twente\\ P.O. Box 217,
7500 AE Enschede, The Netherlands}
\author{Darryl D. Holm}
\address{Theoretical Division and Center for Nonlinear Studies, Los
Alamos National Laboratory\\
{\mbox{MS B284 Los Alamos,}} NM 87545, USA}

\maketitle
\begin{abstract}
A new formulation is derived for the commutator-errors in large-eddy 
simulation of incompressible flow. These commutator-errors arise from
the application of non-uniform filters to the Navier-Stokes 
equations. As a consequence, the filtered velocity field is no longer
solenoidal. The order of magnitude of the commutator-errors is 
compared with the divergence of the turbulent stresses. This shows 
that one
can not reduce the size of the commutator-errors independently of the
turbulent stress terms by some judicious construction of the filter
operator. Similarity modeling for the commutator-errors is presented,
including an extension of Bardina's approach and the application of Leray
regularization. The performance of the commutator-error parameterization is
illustrated with the one-dimensional Burgers equation. For large
filter-width variations the Leray approach is shown to capture the filtered
flow with better accuracy than is possible with Bardina's approach.
\end{abstract}


\newpage

\noindent
{\bf{Corresponding author:}} Bernard J. Geurts

\noindent
{\bf{e-mail:}} b.j.geurts@math.utwente.nl

\noindent
{\bf{tel:}} +31-53-489-4125

\noindent
{\bf{fax:}} +31-53-489-4833


\newpage

The desire to extend large-eddy simulation (LES) to complex flows 
generally implies that one is confronted with strongly varying 
turbulence
intensities within the flow-domain and also as a function of time. In 
certain regions of the flow a nearly laminar flow may arise while a
lively, fine-scale turbulent flow can be present simultaneously in 
another region. In the filtering approach this can be accommodated 
using a
filter operator with non-uniform filter-width that may depend on both 
space and time. The use of such filters, however, further complicates
the subgrid closure problem through the appearance of additional 
commutator-errors \cite{ghosalaiaa}. We will formulate a systematic
modeling of the dynamics of these contributions.

Distinguishing between which flow-features are `subgrid' and which 
are `resolved', depends on the local filter-width $\Delta$. Spatial 
and
temporal variations in
$\Delta$ therefore imply additional energy transfer mechanisms among 
the scales in the flow, {besides} the well-known energy-transfer due 
to
the quadratic nonlinearity in the Navier-Stokes equations. If a flow 
structure propagates from a region of small filter-width into a region
with strongly increased filter-width, it would appear as if part of 
this structure would turn from a `resolved' to a `subgrid' feature,
merely by translation. The reverse can also be imagined, leading to 
the apparent emergence of resolved structures from a collection of
subgrid scales. This suggests additional sources of local energy 
drain or backscatter, depending on the specific local filter-width
variations in the direction of the instantaneous local flow which 
require explicit parameterization.

The traditional use of convolution filters in LES necessarily implies 
that the width of the filter is constant. However, efficient extension
of the LES approach to turbulent flows in complex geometries or to 
cases with strong spatial variation of turbulence intensities, calls 
for
the introduction of non-uniform filter-widths. The starting point is 
an extended filter $L$ which, in one spatial dimension, is defined by
\begin{equation}
\label{nonuniform-filter}
{\ovl{u}}(x,t)=L(u)(x,t)=\int_{x-\Delta_{-}(x,t)}^{x+\Delta_{+}(x,t)} 
\frac{H(x,\xi)}{\Delta(x,t)} u(\xi,t)~d\xi
\end{equation}
where $\Delta_{\pm}$ denote the upper and lower bounds and 
$\Delta=\Delta_{+}+\Delta_{-}$. In complex flow geometries the 
variations in
turbulence intensities poses different local requirements on the 
amount of detail with which the flow should be represented. Such a
situation can be formulated by allowing a non-uniform filter-width as 
given in (\ref{nonuniform-filter}). The application of such filters
gives rise to a number of extra closure terms {in addition to} the 
well-known turbulent stresses.

If one applies the filter (\ref{nonuniform-filter}) to the 
incompressible flow equations, commutator-errors arise since
${\ovl{\pr_{x}u}} \neq \pr_{x}\ub$ or, written differently, 
$L(\pr_{x}u) -\pr_{x}(L(u))=[L,\pr_{x}](u) \neq 0$ in terms of the 
commutator of
$L$ and the derivative operator $\pr_{x}$. For the filtered 
continuity equation we find
\begin{equation}
\label{fildens}
\pr_{j} \baru _{j}= -[L,\pr_{j}](u_{j})
\end{equation}
Hence, the filtered continuity equation is no longer 
in local conservation form and variations in the filter-width imply 
that $\baru_{j}$ is
not solenoidal. Filtering the Navier-Stokes 
equations in the same way 
yields
\begin{eqnarray}
\label{filnavsto}
\pr_{t} \baru_{i} 
&+&\pr_{j}(\baru_{i}\baru_{j})+\pr_{i}{\ovl p} - \frac{1}{Re} 
\pr_{jj} \baru _{i}
=- \Big ([L,\pr_{t}](u_{i})
\nonumber \\
&+& 
\pr_{j}([L,S](u_{i},u_{j})) + [L,\pr_{j}](S(u_{i},u_{j})) \nonumber 
\\
&+&[L,\pr_{i}](p) - \frac{1}{Re}[L,\pr_{jj}](u_{i}) 
\Big)
\end{eqnarray}
We observe that commutators emerge of filtering 
and the product operator $S(f,g)=fg$ as well as commutators of 
filtering and first and
second order partial derivatives. Filtering a 
linear term such as $\pr_{t} u_{i}$ gives rise to a `mean-flow' term 
$\pr_{t}\ub_{i}$ and a
corresponding commutator-error 
$[L,\pr_{t}](u_{i})$. Filtering the nonlinear convective terms leads 
to the divergence of the turbulent
stress tensor 
$\tau_{ij}={\ovl{\baru_{i}\baru_{j}}}-\baru_{i}\baru_{j} 
=[L,S](u_{i},u_{j})$ and an associated 
commutator-error
$[L,\pr_{j}](S(u_{i},u_{j}))$. The local 
conservation form of the Navier-Stokes equations is no longer 
maintained, in the same way as
observed in (\ref{fildens}).

The 
commutators in (\ref{fildens}) and (\ref{filnavsto}) satisfy 
algebraic identities. If we consider any two filters $L_{1}$ and 
$L_{2}$
then
\begin{equation}
[ L_{1} L_{2},S]=[L_{1},S]L_{2}+L_{1} 
[L_{2},S]
\label{germanoidentity}
\end{equation}
which is known as 
Germano's identity \cite{germano}. Likewise,
\begin{equation}
[ 
L_{1},[L_{2},S]]+[L_{2},[S,L_{1}]]+[S,[L_{1},L_{2}]]=0
\label{jacobi}
\end{equation}
which 
is interpreted as Jacobi's identity. These identities are also 
satisfied by $[L,\pr_{t}]$ and $[L,\pr_{j}]$ which shows that 
the
structure of the LES closure problem is closely related to the 
Poisson-bracket in classical mechanics. In that context Germano's 
identity is
known as Leibniz' rule. The identities 
(\ref{germanoidentity}) and (\ref{jacobi}) can be used to guide 
(dynamic) subgrid modeling of the
central commutators.

Filtering the 
incompressible flow equations gives rise to an `LES-template' in 
which the `Navier-Stokes' operator on the left hand side 
of
(\ref{filnavsto}) acts on the filtered solution $\{ \ub_{i},\pb 
\}$. In addition, a number of unclosed terms arises of which only 
the
parameterization of the turbulent stresses attracted considerable 
attention in literature. However, the subgrid modeling problem 
associated
with non-convolution filters entails various additional 
commutator-errors. These terms require explicit modeling in case the 
spatial and
temporal variations of the filter-width are sufficiently 
large. For steady filter-width distributions, to which we restrict 
ourselves here,
the magnitude of these contributions can be 
quantified in terms of ${\ovl{\bf{u}}} \cdot \nabla 
\Delta=\ub_{j}\pr_{j}\Delta$.

The dynamic effects of the 
commutator-errors have been considered unimportant by some authors, 
provided a suitable class of filters would be
adopted. In 
\cite{vdven} such a class of filters was constructed and the 
commutator-errors corresponding to these filters could be made
of 
high order in $\Delta$. Likewise, \cite{vasilyev} considers the 
commutator-errors to be of minor importance in case high order 
filters
are used. Although it is correct that the commutator-errors 
can be made small with the proper filter, one has to realize that 
with the same
filter the divergence of the turbulent stress tensor is 
also reduced to the same order in $\Delta$. The use of higher order 
filters would
hence only imply a gradual convergence to the 
unfiltered Navier-Stokes equations. It is not possible to reduce the 
size of the
commutator-errors {\it independently} of 
$\pr_{j}([L,S](u_{i},u_{j}))$ merely by constructing suitable 
filters. The subgrid modeling of the
dynamic significance of the 
commutator-errors therefore remains a largely open problem.

In order to establish the importance of the commutator-errors 
relative to the turbulent stress contributions we introduce general 
$N$-th
order filters by requiring $L(x^{m})=x^{m}$ for $m=0,~1,\ldots,~N-1$ 
\cite{vasilyev}. Application of such a filter yields:
\[
{\ovl{u}} (x)-u(x)=\sum_{m \geq N} \left( \Delta^{m}(x)M_{m}(x) 
\right) u^{(m)}(x)
\]
where $u^{(m)}$ denotes the $m$-th derivative and $M_{m}(x)$ is 
related to the $m$-th moment of $L$. To leading order
${\ovl{u}} -u \sim \Delta^{N}$. For the commutator-error we find
\[
[L,\pr_{x}](u)=-\sum_{m \geq N} \left( \Delta^{m}M_{m} \right)' u^{(m)}
\]
where the prime denotes differentiation with respect to $x$. The 
commutator $[L,S](u)$ can be expressed as:
\[
[L,S](u)=\sum_{m \geq N} (\Delta^{m}M_{m}) \left( (u^{2})^{(m)}-2u 
u^{(m)} \right) - ({\ovl{u}} -u )^{2}
\]
The scaling of the turbulent stresses with $\Delta^{N}$ is readily 
verified for $N > 1$. In case $N=1$ the commutator scales with
$\Delta^{2}$ since $(u^{2})'=2uu'$. Combining these expressions one 
may obtain the leading order behavior of the flux terms for symmetric
filters as:
\begin{eqnarray}
[L,\pr_{x}](S(u))&\approx& A(x)\left(\Delta^{N-1} \Delta ' M_{N} 
\right) + B(x) \left( \Delta^{N} M_{N}' \right) \nonumber \\
\pr_{x}([L,S](u))&\approx& a(x)\left(\Delta^{N-1} \Delta ' M_{N} 
\right) + b(x) \left( \Delta^{N} M_{N}' \right) \nonumber \\
   &+& c(x)\left(\Delta^{N}M_{N} \right)
\end{eqnarray}
where $A$, $B$, $a$, $b$ and $c$ are smooth, bounded functions which 
contain combinations of derivatives of the solution $u$. For
non-uniform filters $[L,\pr_{x}](S) \sim \Delta' \Delta^{N-1}$ which 
is the same leading order behavior as for $\pr_{x}[L,S]$. Hence, it is
not possible to remove only the commutators $[L,\pr_{x}]$ by a 
careful construction of the filter \cite{vdven}. In fact, all filters 
that
would reduce $[L,\pr_{x}]$ are of higher order and consequently will 
also reduce $[L,S]$ with the same order.

One may use a Fourier-mode analysis to relate the dynamic 
significance of the commutator-errors to variations in $\Delta$ and 
the wavenumber
$k$ of the mode. For second order filters such as the top-hat or 
Gaussian filter one has
\be
\frac{\| [L,\pr_{x}](S(u)) \|}{\| \pr_{x}([L,S](u))\|} \sim 
\frac{|\Delta'/\Delta|}{|k|}
\en
where $\| \cdot \|$ denotes the $L_{2}$-norm. This shows that if 
$|\Delta'| \ll |k\Delta|$ then filter-width non-uniformity can be
disregarded and it should be sufficient to model only $\tau_{ij}$. 
This shows that only strongly bounded variations in the filter-width 
will
reduce the size of the commutator-errors significantly while keeping 
the magnitude of the turbulent stresses unaffected. If, however, for
efficiency reasons or due to, e.g., wall-proximity, a sufficiently 
smooth variation of $\Delta$ is not possible, one has to resort to 
direct
modeling of the commutator-errors.

In the absence of a comprehensive theory of turbulence and its 
non-uniform spatial and temporal representations, the modeling of the
turbulent stresses and the commutator-errors relies to some degree on 
limited empirical knowledge. Here we restrict ourselves to similarity
modeling and consider two different approaches. Specifically, we will 
extend Bardina's approach~\cite{bardina} to include commutator-errors
and we derive the implied subgrid models arising from Leray 
regularization~\cite{leray}.

Bardina's similarity model for the turbulent stress tensor arises 
from applying the definition of $\tau_{ij}$ to $\ub_{i}$, i.e.,
\be
\tau_{ij} \rightarrow 
m_{ij}^{B}=[L,S](\ub_{i},\ub_{j})={\ovl{\ub_{i}\ub_{j}}}-{\ovl{\ub}}_{i}{\ovl{\ub}}_{j}
\en
Extending this idea to the commutator-error suggests the following 
parameterization: $[L,\pr_{j}](u_{i}u_{j}) \rightarrow
[L,\pr_{j}](\ub_{i}\ub_{j})$. In \cite{geurtscrete} this model showed 
a high correlation for turbulent boundary layer flow which partially
substantiates this approach.

Recently, the Leray regularization 
principle \cite{leray} was revived in the context of LES 
\cite{gh2002}. In this approach the convective
fluxes are replaced by 
$\ub_{j}\pr_{j}u_{i}$, i.e., the solution $\uu$ is convected with a 
smoothed velocity $\uub$. The governing Leray
equations can be 
written as
\cite{leray}
\begin{equation}
\pr_{j} 
u_{j}=0~~;~~\pr_{t}u_{i}+\ub_{j}\pr_{j}u_{i}+\pr_{i}p-\frac{1}{Re}\pr_{jj}u_{i}=0
\label{leray1}
\end{equation}
This 
formulation can be written in terms of $\{ \ub_{i},\pb \}$ in case we 
assume a (formal) inverse $L^{-1}$ of $L$, 
i.e.,
$u_{j}=L^{-1}(\ub_{j})$. After some calculation one obtains the 
filtered continuity equation (\ref{fildens}) and the filtered 
momentum
equation as
\begin{eqnarray}
\pr_{t} \baru 
_{i}&+&\pr_{j}(\baru_{i}\baru_{j})+\pr_{i}{\ovl p} - \frac{1}{Re} 
\pr_{jj} \baru _{i} =
- \Big([L,\pr_{t}](u_{i}) \nonumber \\
&+& 
\Big\{\pr_{j}(m_{ij}^{L})+{\ovl{u_{i}\pr_{j}\baru_{j}}}\Big\} + 
[L,\pr_{j}](S(u_{i},\baru_{j})) \nonumber \\
&+& [L,\pr_{i}](p) - 
\frac{1}{Re}[L,\pr_{jj}](u_{i}) 
\Big)
\label{otherleray}
\end{eqnarray}
The divergence of the 
turbulent stress tensor in (\ref{filnavsto}) is represented in terms 
of the asymmetric, filtered similarity-type
Leray model $m_{ij}^{L}$ 
and an additional term associated with the divergence of the filtered 
velocity field: 
\begin{equation}
\pr_{j}\tau_{ij} \rightarrow 
\pr_{ij}( 
m_{ij}^{L})+{\ovl{u_{i}\pr_{j}\baru_{j}}}
\label{basic_leray}
\end{equation}
where 
$m_{ij}^{L}=\ovl{\ub_{j}u_{i}} -\ub_{j}\ub_{i}$ \cite{gh2002} and the 
commutator-error is expressed as
$[L,\pr_{j}](u_{i}u_{j}) \rightarrow 
[L,\pr_{j}](\ub_{j}u_{i})$. The other commutator-errors are identical 
to those in (\ref{filnavsto})
with the understanding that in actual 
simulations every occurrence of an unfiltered flow-variable implies 
the application of $L^{-1}$ to the
available field. The Leray model 
was shown to provide good predictions of three-dimensional turbulent 
mixing at arbitrarily high Reynolds
number using a uniform filter 
\cite{gh2002}.

To assess the effects of the commutator-errors and 
determine the quality of the Bardina and Leray modeling we consider 
the one-dimensional
Burgers equation. This provides a model-system 
which has the same basic structure under filtering as expressed in 
(\ref{filnavsto}). All
relevant commutators appear in the filtered 
Burgers equation. The initial solution is a Gaussian profile which 
rapidly develops into the
well-known `ramp-cliff' structure. We use 
$Re=500$ to obtain a sharply localized cliff region, and apply 
periodic boundary conditions.
Explicit time-integration, restricted 
by stability time-steps, and second order accurate spatial 
discretization are adopted. To avoid
numerical errors we use high 
spatial resolution, typically with $N=2048$ intervals. Explicit 
filtering is done with trapezoidal quadrature
applied to the top-hat 
filter.

We consider a non-uniform grid with grid-spacing 
$h_{i}=(\ell/N)(1+g_{i})$ where $\ell$ is the length of the domain. 
The grid is chosen to
be non-uniform only in an interval around 
$i=N/2$ in computational space. For the illustrations we 
use
\[
g_{i}=A\sin \Big(2\pi\frac{(i-N/2)}{(N(m-2q)/m)} 
\Big)~~~;~~\frac{qN}{m} \leq i \leq \frac{(m-q)N}{m}
\]
and 0 
otherwise. Since $\sum g_{i}=0$ this grid preserves the end-points. 
The parameters $q$ and $m$ control the region where the grid 
is
non-uniform. Here, we use $q=3$ and $m=8$. The local filter-width 
$\Delta_{i}=x_{i+n}-x_{i-n}$ where $n$ is chosen such that 
$\Delta=\ell/D$
with $D=8$ or $16$ in the uniform regions of the 
grid. With $N=2048$ this implies $n=128$ or $64$ respectively. The 
parameter $A < 1$
controls the ratio between largest and smallest 
intervals $(1+A)/(1-A)$.

In figure~\ref{fluxes} we collected the 
contributions to the total convective flux for a representative 
uniform and non-uniform case. We
decomposed the convective flux 
as
\[
{\ovl{\pr_{x}(u^{2})}}=\pr_{x}({\ovl{u}}^{2})+\pr_{x}({\ovl{u^{2}}}-{\ovl{u}}^{2})+\{{\ovl{\pr_{x}(u^{2})}}-
\pr_{x}({\ovl{u^{2}}}) \}
\]
identifying on the right hand side the 
`mean' flux, the `SGS-flux' and the `commutator-flux' respectively. 
In figure~\ref{fluxes} the
solution and the filtered solution are 
included displaying the `ramp-cliff' structure. The total flux 
in
figure~\ref{fluxes}(a) is piecewise linear and the SGS flux is 
localized in the cliff-region where filtering is effective. 
In
figure~\ref{fluxes}(b) there is a significant distortion of the 
filtered solution due to the filter-width non-uniformity. Two 
characteristic
contributions due to the commutator-error arise. On 
the `ramp side', the non-uniform filter-width near $x=-3$ strongly 
influences the mean
flux. The commutator-error compensates for this 
such that the total flux remains nearly linear in $x$. Within the 
`cliff-region' the
commutator-flux is comparable to the SGS-flux.

In 
figure~\ref{xlocminmax}(a) we show the locations of the front and 
back of the ramp-cliff solution as a function of time. These 
locations
are defined where $|u|$ equals $\varepsilon \max(|u|)$ with 
$\varepsilon=0.05$. Comparing filtered Burgers results with 
predictions from the
Leray and Bardina parameterizations, the Leray 
results are more accurate. This was confirmed by considering the 
minimal and maximal values
of $u$ which are also better predicted by 
the Leray model. The $L_{2}$-norm of the fluxes for these cases show 
that the commutator-flux is
about 1/3-1/2 the value of the SGS-flux. 
In cases with smaller grid non-uniformities the direct Leray modeling 
still enhances the accuracy
of the predictions notably. The Leray 
model is considerably less expensive than the Bardina model and it 
also better preserves qualitative
properties of the filtered Burgers 
solution cf. figure~\ref{xlocminmax}(b). The Bardina parameterization 
creates additional structure in the
solution, which is not present in 
the filtered Burgers result.

The commutator-errors have been 
expressed as commutators of filtering and partial derivatives. The 
magnitude of the commutator-errors can
not be reduced independently 
of the SGS-fluxes. Instead, for sufficiently large grid 
non-uniformities direct modeling is needed. For the
one-dimensional 
Burgers equation the Leray parameterization combines computational 
efficiency with high accuracy. This motivates the use of
the Leray 
commutator model in more complex flows. The a priori specification of 
the spatial filter-width variations is not generally
possible for 
complex cases. Therefore, the local filter-width needs to be related 
to the resolved solution to facilitate a dynamic response
of local 
filtering and the evolving flow. This is subject on ongoing 
research.


\section*{Acknowledgment}

BJG would 
like to acknowledge support from the Los Alamos Turbulence Working 
Group and Center for Nonlinear Studies 
(2002).


\newpage

\begin{figure}[htb]

\caption{Snapshot 
of the solution (multiplied by 1/2) (solid) and filtered solution 
(solid; markers $o$). Convective flux: total (dots),
mean 
(dash-dotted), turbulent stress (dashed), commutator-error (solid 
with $*$). In (a) we use $\Delta=\ell/16$ and in (b) the 
non-uniform
case with $A=1/2$ is shown. Underneath in (b), the 
grid-spacing (minus 0.2) as a function of $x$ is presented. 
}

\end{figure}

\begin{figure}[htb]

\caption{Location of the head 
of the cliff (upper curves) and the tail of the ramp (lower curves) 
in (a) and in (b) snapshot of the filtered
solution: filtered Burgers 
(solid), Leray (dashed) and Bardina (dash-dotted) for 
$A=0.85$}

\end{figure}

\newpage

\setcounter{figure}{0}

\begin{figure}[htb]

\centerline{
\psfig{figure=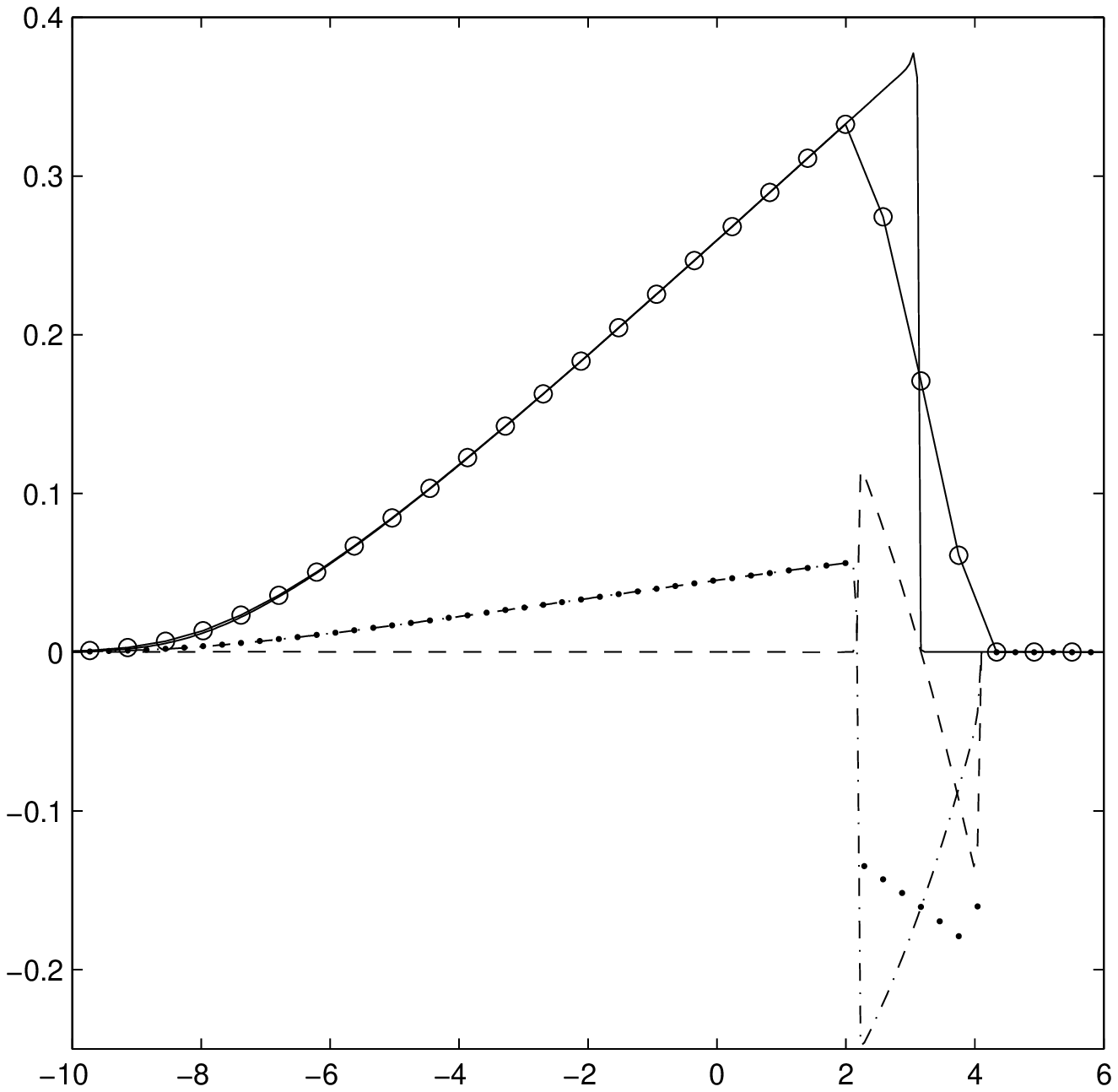,width=0.6\textwidth}(a)
}

\hspace*{0.6\textwidth} 
{\Large{$x$}}

\vspace*{3mm}

\centerline{
\psfig{figure=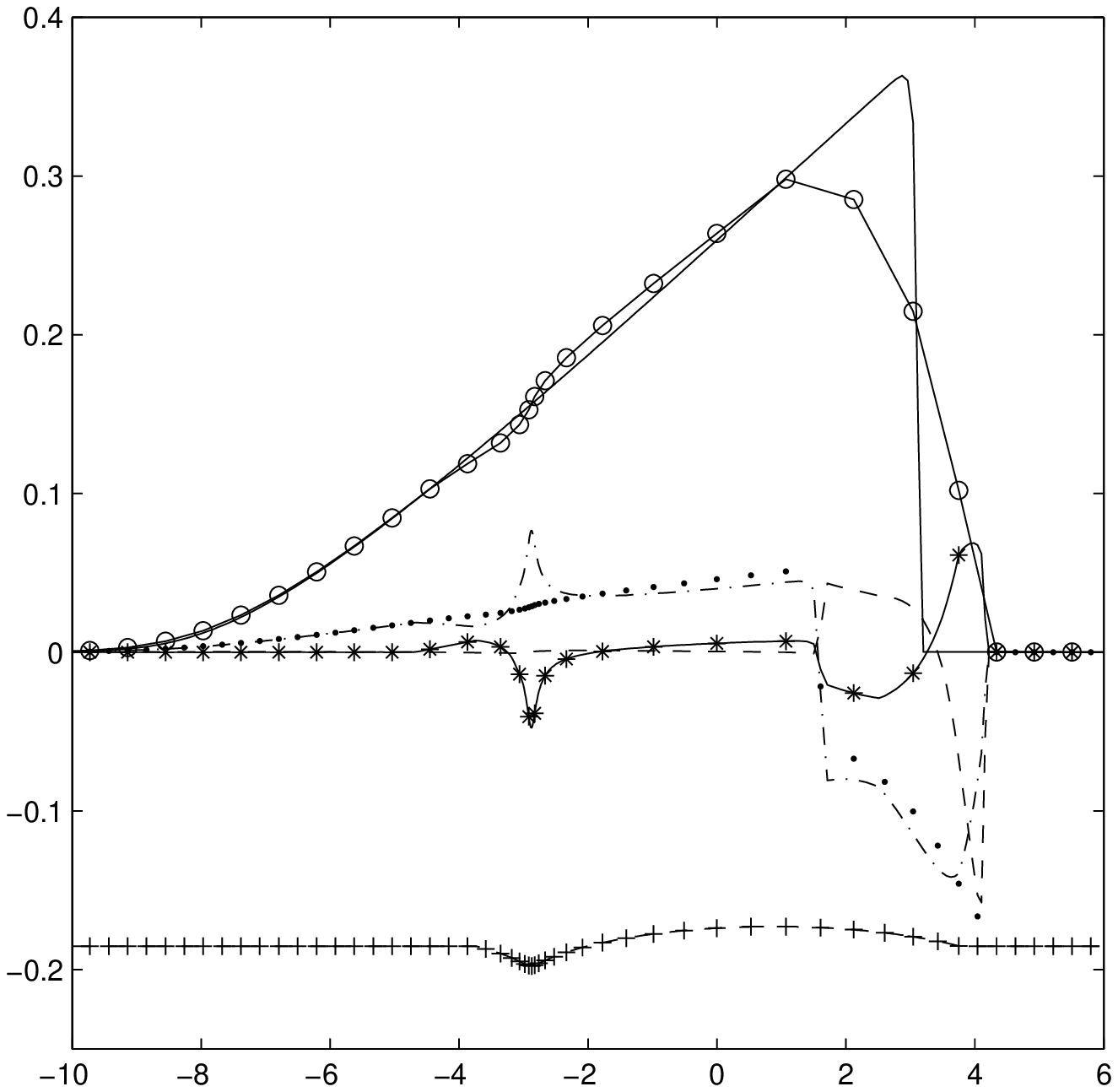,width=0.6\textwidth}(b)
}

\hspace*{0.6\textwidth} 
{\Large{$x$}}

\vspace*{10mm}

\caption{Geurts, Holm: 
Phys.Fluids}

\label{fluxes}
\end{figure}

\newpage

\begin{figure}[htb]

\centerline{
\psfig{figure=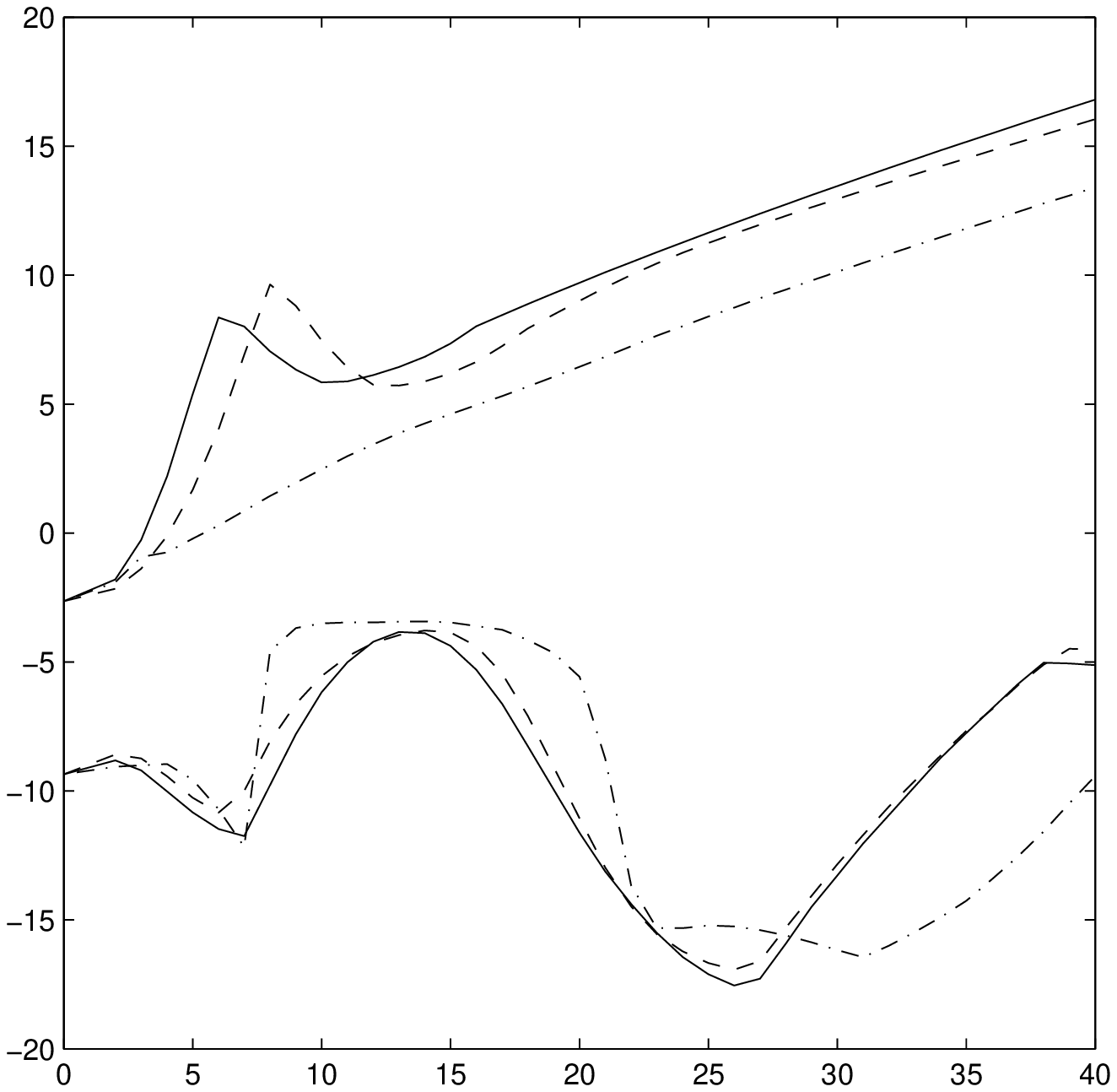,width=0.6\textwidth}(a)
}

\vspace*{-0.4\textwidth}

\hspace*{12.5mm} 
{\Large{$x_{head}$}}

\vspace*{0.15\textwidth}

\hspace*{12.5mm} 
{\Large{$x_{tail}$}}

\vspace*{0.125\textwidth}

\hspace*{0.6\textwidth} 
{\Large{$t$}}

\vspace*{3mm}

\centerline{
\psfig{figure=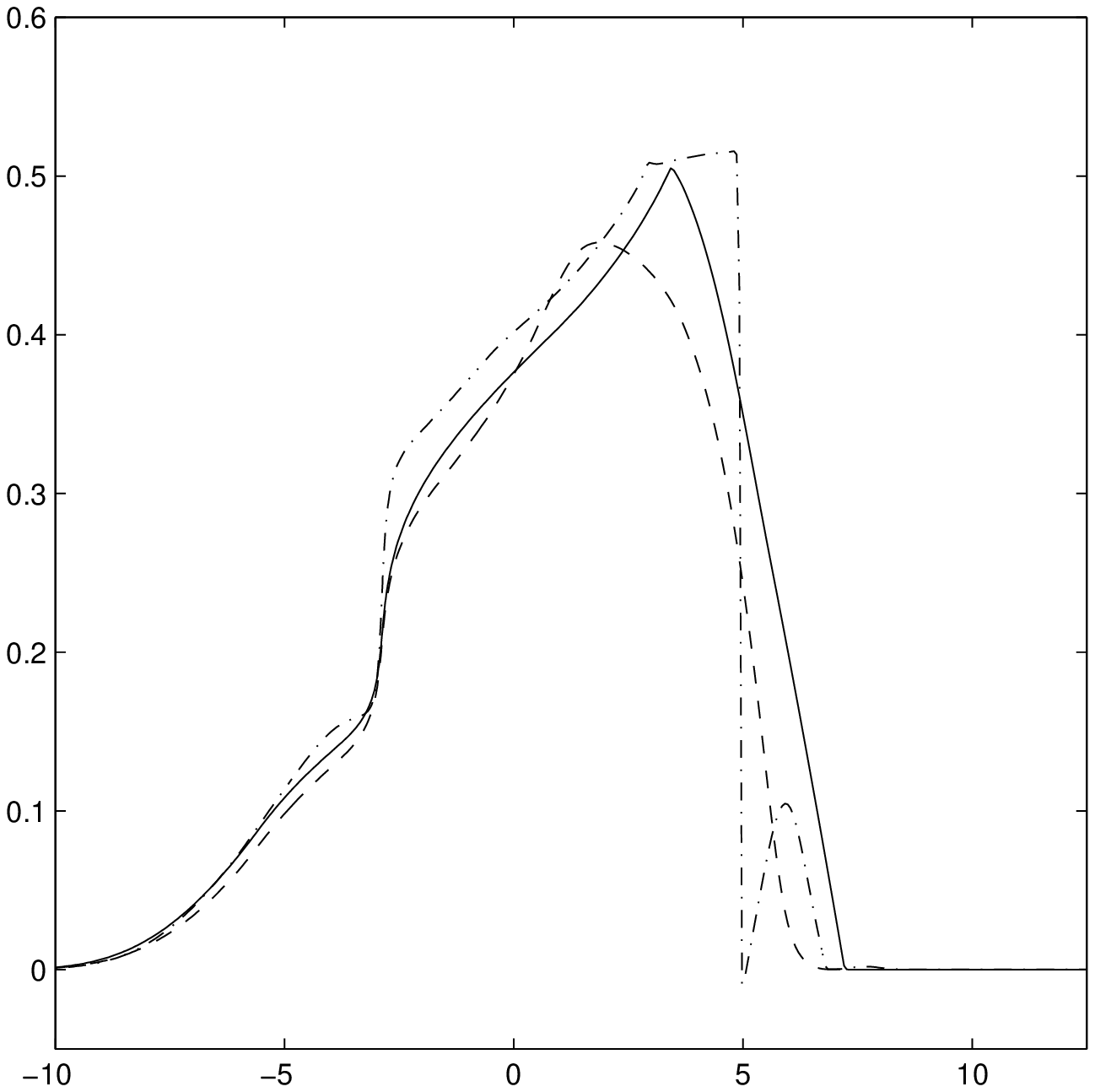,width=0.6\textwidth}(b)
}

\vspace*{-0.55\textwidth}

\hspace*{15mm} 
{\Large{$\ub$}}

\vspace*{0.475\textwidth}

\hspace*{0.6\textwidth} 
{\Large{$x$}}

\vspace*{5mm}

\caption{Geurts, Holm: 
Phys.Fluids}

\label{xlocminmax}

\end{figure}


\end{document}